\documentclass[twocolumn,superscriptaddress,showpacs,preprintnumbers,amsmath,amssymb]{revtex4}
\usepackage{graphicx}% Include figure files
\usepackage{dcolumn}% Align table columns on decimal point
\usepackage{bm}% bold mat

\usepackage{amsmath,amssymb}

\newcommand{\bk}{{{\bf{k}}}}

\newcommand{\br}{{{\bf{r}}}}

\newcommand{\bS}{{\bf S}}

\newcommand{\dg}{{\dagger}}

\newcommand{\nn}{\nonumber}          
\newcommand{\la}{\langle}          
\newcommand{\ra}{\rangle}          
\newcommand{\pbar}{\bar{p}}          %{\overbar{p}}          
\newcommand{\bea}{\begin{eqnarray}}          
\newcommand{\eea}{\end{eqnarray}}

\begin{document}

\title{Plaquette-RVB state in the Frustrated Honeycomb Antiferromagnet}

\author{R. Ganesh}
\email[electronic address: ]{g.ramachandran@ifw-dresden.de}
\affiliation{Institute for Theoretical Solid State Physics, IFW Dresden,
PF 270116, 01171 Dresden, Germany}
\author{Satoshi Nishimoto}
\affiliation{Institute for Theoretical Solid State Physics, IFW Dresden,
PF 270116, 01171 Dresden, Germany}
\author{Jeroen van den Brink}
\affiliation{Institute for Theoretical Solid State Physics, IFW Dresden,
PF 270116, 01171 Dresden, Germany}
\affiliation{Department of Physics, Technical University, D-1062 Dresden, Germany}
\date{\today}

\begin{abstract}
We study the proposed plaquette-RVB (pRVB) state in the honeycomb lattice $J_1-J_2$ model with frustration arising from next-nearest neighbour interactions. Starting with the limit of decoupled hexagons, we develop a plaquette operator approach to describe the pRVB state and its low energy excitations. Our calculation clarifies that the putative pRVB state necessarily has f-wave symmetry - the plaquette wavefunction is an antisymmetric combination of the Kekul\'e structures. We estimate the plaquette ordering amplitude, ground state energy and spin gap as a function of $J_2/J_1$. 
The pRVB state is most stable around $J_2/J_1 \sim 0.25$. We identify the wavevectors of the lowest triplet excitations, which can be verified using exact diagonalization or DMRG studies. When $J_2$ is reduced, we can have either a deconfined Quantum Phase Transition (QPT) or a first-order transition into a N\'eel state. When $J_2$ is increased, we surmise that the system will undergo a first order phase transition into a state which breaks lattice rotational symmetry. 
\end{abstract}

\maketitle

\section{Introduction}
\label{sec.intro}
Magnetic systems with frustrating interactions serve as an excellent route to generating novel quantum states with various degrees of entanglement. They often give rise to many body ground states in which many constituents become entangled, from which new collective degrees of freedom emerge. Examples include Valence Bond Solid (VBS) phases\cite{Sawtooth}, decoupled chains\cite{Starykh} and spin liquids. At each level of entanglement, a new effective theory is needed to describe the low energy properties of the system. In this light, we study a plaquette-ordered phase proposed as a ground state for the honeycomb lattice spin-1/2 $J_1-J_2$ model, with frustration arising from next-nearest neighbour couplings. This plaquette-RVB (pRVB) phase comprises of entangled hexagons, arranged in a $\sqrt{3}\times\sqrt{3}$ pattern. 

Recently, the J$_1$-J$_2$ model on the honeycomb lattice has generated tremendous interest as an effective model for the intermediate-U physics of the Hubbard model on the honeycomb lattice\cite{Meng}. It may also be relevant to the material Bi$_3$Mn$_4$O$_{12}$(NO$_3$) which shows no long range order down to the lowest temperatures\cite{BMNO,Sheng,Kawamura}. 
%Recently, there has been tremendous interest in this $J_1-J_2$ model due to their relevance to the intermediate-U physics of the honeycomb lattice Hubbard model and to the material Bi$_3$Mn$_4$O$_{12}$(NO$_3$). The simplest model, with frustration arising from next-nearest neighbour $J_2$ interactions, has been extensivley studied recently. 
Being the simplest model of frustration on the honeycomb lattice, the $J_1-J_2$ model has been extensively studied. In the semiclassical limit of large-S, this model is very well understood: for small $J_2$, the ground state has N\'eel order. At a critical value $J_2 = 1/6$ (we set J$_1$ = 1), there is a Lifshitz transition into a spiral state with extensive ground state degeneracy. Order-by-disorder effects mediated by quantum/thermal fluctuations select three spiral states which are related by lattice rotations. The ground state thus breaks lattice rotational symmetry. 
By contrast, the extreme quantum limit of $S=1/2$ is not well understood. In spite of many recent studies, the phase diagram has not yet been conclusively established.
In particular, many studies point to an interesting intermediate phase for $0.2\lesssim$J$_2 \lesssim$0.4 which interpolates between N\'eel order and a state which breaks lattice rotational symmetry.
%
%For small values of $J_2$, the ground state is known to be N\'eel ordered. For $J_2 \gtrsim0.4$, the ground state may be a valence bond solid or a magnetically ordered state which breaks lattice rotational symmetry. Many studies point to an interesting intermediate phase which interpolates between N\'eel order and the large-$J_2$ phase. 

While all studies in the literature agree upon the existence of an intermediate phase, its precise nature has not been established. Some proposals indicate that the intermediate state may be a $Z_2$ spin liquid\cite{FaWang,Ran,Clark}. Others give support to a pRVB state which breaks translational symmetry\cite{Jafari,Albuquerque,cluster}. A functional Renormalization Group study shows weak tendency to pRVB order\cite{Thomale}, while a recent variational calculation does not find plaquette order\cite{Boninsegni}. 
In the closely related $J_1-J_2-J_3$ model, coupled cluster calculations support pRVB order\cite{Richter}. 
In the light of such conflicting studies, we study the proposed pRVB state and suggest more precise tests to confirm its existence. It is worthwhile to devise such tests as most of the evidence for a pRVB ground state comes from calculations restricted to small system sizes. In addition, our study sheds light on the properties of this putative state, its low energy excitations and the nature of phase transitions into and out of this phase. 

In this paper, we build an effective theory for the candidate pRVB state using a \textit{plaquette-operator} approach. We estimate its ground state energy and spin gap and determine the nature of low-energy excitations. Our goals are threefold: (i) to understand the nature of the pRVB state, (ii) to clarify the nature of phase transitions into and out of this state, and (ii) to suggest tests which can be carried out using numerical techniques such as the Density Matrix Renormalization Group (DMRG). Such tests might conclusively reveal whether the pRVB state indeed arises in the J$_1$-J$_2$ model.

This paper is organized as follows. In Sec. \ref{sec.singlehexagon}, we begin with the S$=$1/2 $J_1-J_2$ model on a single hexagon. We classify all its eigenstates as a function of $J_2/J_1$. Sec. \ref{sec.poperatortheory} introduces the plaquette operator representation and lays out the formalism sequentially - by first considering two coupled hexagons and then tiling hexagons to form a honeycomb lattice. We argue that the pRVB state has f-wave character, rather than s-wave. We arrive at a simplified effective Hamiltonian that only involves triplet excitations.
Sec. \ref{sec.results} gives the results of the pRVB mean-field theory with spin-gap, ground state energy and pRVB amplitude. Finally, Sec.\ref{sec.discussion} discusses the nature of phase transitions into and out of this phase and implications for the $J_1-J_2$ model.

\begin{figure}
\includegraphics[width=2.5in]{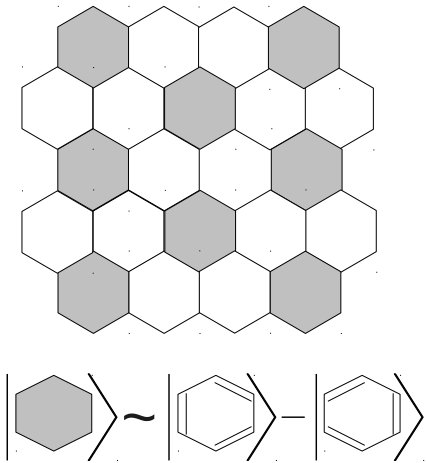}
\caption{The proposed pRVB state on the honeycomb lattice. The shaded hexagons are in an anti-symmetric combination of the two singlet covers}.
\label{fig.prvb}
\end{figure}

\section{The single hexagon problem}
\label{sec.singlehexagon}

\begin{figure}
\includegraphics[width=3.5in]{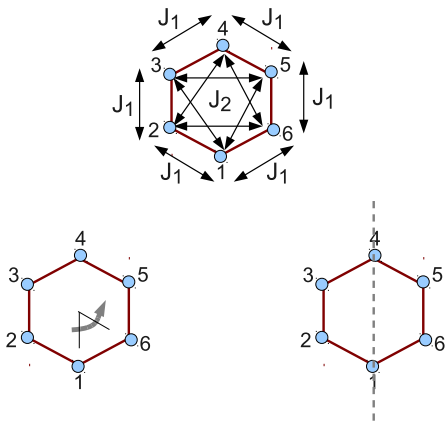}
\caption{Top: Exchange couplings in the single hexagon Hamiltonian. The following operations commute with the Hamiltonian:  (bottom left) rotation by 60 $^\circ$, and (bottom right) inversion about an axis passing through opposite sites.}
\label{fig.singhxgn}
\end{figure}

On a single hexagon, the spin-1/2 J$_1$-J$_2$ model is easily diagonalized, with a Hilbert space of $64$ $(=2^6)$ states. The Hamiltonian is given by (we set J$_1$=1)
\bea
H_{single-hex} = \sum_{\la ij \ra}\bS_{i}\cdot\bS_{j} + J_2\sum_{\la\la ij\ra\ra}\bS_{i}\cdot\bS_{j}.
\eea
We take the exchange couplings J$_1$ and J$_2$ to be positive. The J$_1$ and J$_2$ bonds are shown in Fig.\ref{fig.singhxgn}. %The Hilbert space has $2^{6} = 64$ states. 

As the Hamiltonian is invariant under global spin rotations, its eigenstates have well-defined magnetic quantum numbers. 
%The Hamiltonian preserves total spin:
%\bea
%[H,\hat{S}_{tot}^z]=[H,\hat{S}_{tot}^2]=0.
%\eea
Diagonalizing the Hamiltonian numerically, we obtain the spectrum consisting of 5 singlets (S$_{tot}$=0), 27 triplets (S$_{tot}$=1), 25 quintets (S$_{tot}$=2) and 7 heptets (S$_{tot}$=3). In the next section, we will argue that only one singlet and the 27 triplet states are relevant for understanding the pRVB phase.

The hamiltonian is invariant under two spatial operations: (i) rotation by 60$^\circ$, and (ii) mirror inversion about a line (see Fig.\ref{fig.singhxgn}). As the two operations do not commute with each other, the eigenstates of the hamiltonian cannot always be chosen to be simultaneously the eigenstates of both. For later convenience, we choose the eigenstates to be simultaneously eigenstates of rotation. Defining $\hat{R}$ to be the rotation operator, we note that $\hat{R}^6$ is the identity operation. Therefore, the eigenvalues of $\hat{R}$ are the sixth roots of unity.

%The inversion operator, $\hat{I}$ does not commute with rotation. Thus, the eigenstates of the Hamiltonian cannot be chosen to be simultaneously the eigenvalues of both $\hat{I}$ and $\hat{R}$. For later convenience, we take the eigenstates to be simultaneously the eigenstates of the rotation operator.
 %Similarly, we denote the inversion operator as $\hat{I}$. $\hat{I}^2$ is again the identity operation, therefore the eigenvalues of $\hat{I}$ are $\pm 1$.

We denote the states obtained by diagonalization as $\vert u \ra$, with $u=1,\ldots,64$ with the energy eigenvalues $\epsilon_u$. Each state is assigned three quantum numbers $(l_u,m_u,r_u)$ where $l_u,\phantom{a}m_u$ are the usual magnetic quantum numbers and $r_u$ is the eigenvalue under rotation. For example, $(l_u,m_u,r_u)=(1,0,-1)$ indicates that the state $\vert u \ra$ is a triplet with $S_z=0$ and it changes sign under rotation by 60$^\circ$. Note that these three quantum numbers do not uniquely identify a state. 

As $J_2$ is varied, some of the the low energy states of the single hexagon problem are:

(i) `+' singlet: this state is the \textit{symmetric} combination of the two Kekul\'e structures proposed for benzene. With quantum numbers $(0,0,1)$, it is invariant under rotation by 60$^\circ$. 

(ii) `-' singlet: this is the ground state for $J_2<0.5$. It is predominantly ($\sim 98.5\%$ - the precise fraction depends weakly on $J_2$) an \textit{antisymmetric} combination of the Kekul\'e structures (see Fig.\ref{fig.prvb}). With quantum numbers $(0,0,-1)$, it is even under inversion and odd under rotation. 
%For small $J_2$ values, when hexagons are tiled to form a two dimensional lattice, we expect to have spontaneous symmetry breaking leading to N\'eel order. At the single hexagon level, this `-' singlet can be thought of as the lowest state of the N\'eel Anderson tower. 

(iii) $t_1$ triplets: these are the triplet states 
%of the N\'eel Anderson tower, 
with the quantum numbers $(1,m,1)$, with $m=-1,0,1$. They are invariant under rotation.

\begin{figure}
\includegraphics[width=2.5in]{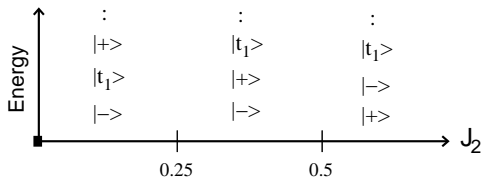}
\caption{Energy spectrum of the single hexagon problem - for brevity, we only show the lowest three states. The ground state is the `-' singlet for $J_2<0.5$. For $J_2>0.5$, the `+' singlet becomes the ground state. For $J_2<0.25$, the first excited states are the $t_1$ triplets. For $J_2>0.25$, the $t_1$ go higher in energy. 
}
\label{fig.singhxgnPD}
\end{figure}

Fig [\ref{fig.singhxgnPD}] shows the low-energy spectrum of the single hexagon problem as a function of $J_2$. We only indicate the ground state and the first two excited states. The `-' singlet is the ground state for $J_2<0.5$. There is a level crossing at $J_2=0.5$ beyond which the `+' singlet becomes the ground state.  

On the full honeycomb lattice, previous studies have suggested a plaquette-ordered ground state for $0.2 \lesssim J_2 \lesssim 0.4$\cite{Jafari,Albuquerque,cluster}. In this parameter regime, the ground state of the \textit{single} hexagon problem is the `-' singlet. We argue that the suggested plaquette-ordered state is composed of a $\sqrt{3}\times\sqrt{3}$ ordering of hexagons in the `-' state as shown in Fig.\ref{fig.prvb}. The choice of the `-' state corresponds to f-wave pRVB order - the plaquette wavefunction changes sign upon rotation by 60$^\circ$. 
In the next section, we describe the plaquette-operator method which provides a consistent scheme to include quantum fluctuations around this state. 

\section{Plaquette operator theory}
\label{sec.poperatortheory}

We obtain an effective theory of the pRVB state in analogy with the bond operator formalism of Sachdev and Bhatt\cite{SachdevBhatt}. Developed in the context of dimer states in spin-1/2 systems, the bond operator method starts from the limit of isolated bonds - the ground state is then composed of disconnected dimers (singlets). A bosonic representation is introduced for the states of each bond in the singlet-triplet basis, with a constraint of unit boson occupancy to preserve the Hilbert space. The dimer state can now be elegantly described as a Bose-condensate of singlet bosons. Moving away from the limit of isolated bonds, inter-bond coupling can be naturally interpreted as four-particle interactions between bosons. Assuming that dimer order is strong, one arrives at a mean-field prescription to treat the interaction terms which gives rise to a quadratic theory of `triplons'. Corrections beyond mean-field theory, viz., interactions between triplons, can be ignored as the triplons are present in dilute concentrations. This theory has been very succesful in obtaining the spectrum of triplon excitations above the dimer state. The dimer-N\'eel phase transition is easily understood as the Bose-condensation of triplons with the bond-operator mean-field theory providing excellent quantitative estimates for the transition point (e.g., see Refs.\cite{Matsushita,Sandvik}).

This approach has been adapted to study plaquette order, albeit in a simple system with four-site plaquettes\cite{Brijeshplaquette,Isaev}. Here, we develop a plaquette operator approach to describe six-site plaquette order in the honeycomb pRVB state. 

Having obtained all eigenstates of the single hexagon problem in the previous section, we now introduce a bosonic representation for these states.
We associate a boson with each of the 64 states in the Hilbert space. We call these `plaquette-bosons'. To preserve the Hilbert space, we impose a constraint of unit boson occupany. On a single hexagon, we may write
\bea
\vert u \ra = b_u^\dg \vert 0 \ra.
\eea
Here, the operator $b_u^\dg$ creates a \textit{u}-boson when it acts upon the vacuum. The vacuum itself is an unphysical state, as it does not respect the unit occupancy constraint. Operators can also be translated into the new representation, e.g., $\vert v \ra \la u \vert \equiv b_v^\dg b_u$. 

Having solved the single hexagon problem exactly, we tile hexagons to form a two-dimensional honeycomb lattice, as shown in Fig. \ref{fig.prvb}. The shaded hexagons serve as sites upon which plaquette-bosons reside. 
As a starting point, we consider the limit of isolated plaquettes. Turning off all inter-plaquette couplings, we obtain a quadratic theory of bosons given by
\begin{eqnarray}
H_{intra-plaq.}  = \sum_{i} \sum_{u} \epsilon_u b_{i,u}^\dg b_{i,u},
\end{eqnarray}
where the index $i$ sums over all shaded plaquettes in Fig.\ref{fig.prvb} - note that the shaded hexagons form a triangular lattice.  The index $u$ sums over the 64 eigenstates of the single hexagon problem. The operator $b_{i,u}^\dg$ creates a $u$ boson at the site (plaquette) $i$. 

The pRVB state can now be identified as a Bose-condensate of the `-' bosons; i.e., we may set 
\bea
b_{i,-} \equiv b_{i,-}^\dg \equiv \pbar,
\eea
where $\pbar$ is the Bose-condensation amplitude. Physically, the quantity $\pbar^2$ gives the probability of finding a shaded plaquette in the `-' state. We take $\pbar$ to be real as its phase is not relevant in the present context. In the isolated plaquette limit, $\pbar^2$ is unity as every hexagon is entirely in the `-' state. 

To preserve the Hilbert space, the total occupancy of bosons per plaquette should be unity. In the spirit of mean-field theory, we enforce this constraint on average using a chemical potential $\mu$.
\bea
H \rightarrow H-\mu \Big[ N\pbar^2  + \sum_{i,u=2,\ldots,64} b_{i,u}^\dg b_{i,u} -N \Big],
\eea
where $N$ is the total number of shaded plaquettes in Fig.\ref{fig.prvb}. Eventually, we will set $\mu$ by demanding
\bea
\frac{\partial \la H \ra}{\partial \mu}=0,
\eea
which will tune the average boson occupancy to unity. 

The next step towards obtaining a consistent theory is to introduce inter-plaquette terms. We deduce the form of these terms by considering two coupled hexagons in the following section. Before giving details of our plaquette operator theory, we illustrate the approach by an analogy to spin wave theory. Consider a spin-S system that has Neel order at zero temperature. To zeroth order, this is described as a classical system of anti-aligned spins with each spin of length S. This is analogous to our isolated plaquette limit with $\pbar=1$ which has perfect plaquette ordering. Standard linear spin wave theory takes into account quantum fluctuations about the N\'eel state and gives the spectrum of low-lying excitations. Spin waves reduce the length of the ordered moment and give rise to zero-point energy corrections. Analogously, our plaquette operator theory captures the low-lying excitations about the pRVB state. These corrections will reduce the value of $\pbar$ and modify the ground state energy in a self-consistent manner.

\subsection{Coupling two hexagons}
\label{ssec.couplingtwo}

\begin{figure}
\includegraphics[width=1.5in]{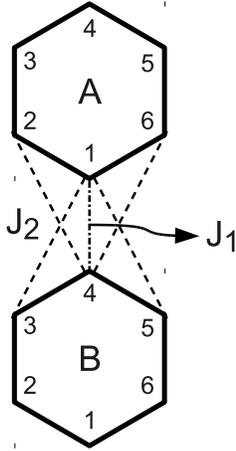}
\caption{Coupling between two `nearest-neighbour' hexagons, A and B. The dotted lines denoted Heisenberg-like exchange terms which couple A and B. 
}
\label{fig.twohexagons}
\end{figure}

Fig.\ref{fig.twohexagons} shows two nearest-neighbour hexagons, A and B, and the five bonds which couple them. The five Heisenberg-like bonds act as a perturbation, which we denote as $H_{AB}$, coupling the single-hexagon states of A and B. In the absence of this perturbation, the total eigenstate is a direct product of single hexagon states centred on A and B - $\vert A_u B_v \ra$, signifying that the A hexagon is in the $\vert u \rangle$ state and B is in the $\vert v \rangle$ state. 

The matrix element of $H_{AB}$ between two direct product states, $\vert A_u B_v \ra$ and $\vert A_p B_q \ra$, is given by  
\bea
\nn \la A_u B_v \vert H_{AB} \vert A_p B_q \ra = \\
\sum_{b} \sum_{\alpha=x,y,z}  J_{b} \la A_{u} B_{v} \vert S_{i_b,A,\alpha} S_{j_b,B,\alpha} \vert A_{p} B_{q} \ra.
\eea
The index $b$ sums over the inter-hexagon bonds which connect site $i_b$ of hexagon A and site $j_b$ of hexagon B (see Fig. \ref{fig.twohexagons}). Each bond has an exchange coupling strength $J_b$ (given by $J_1$ for one bond and $J_2$ for four other bonds). 

This quantity reduces to a product of matrix elements of single-hexagon operators,
\bea
\nn\la A_u B_v \vert H_{AB} \vert A_p B_q \ra = \sum_{b} \sum_{\alpha=x,y,z}  J_{b} \la A_{u} \vert S_{i_b,A,\alpha} \vert A_p \ra \\
\times \la B_{v} \vert  S_{j_b,B,\alpha} \vert B_{q} \ra.
\label{eq.ABcoupling}
\eea

Using Eq.\ref{eq.ABcoupling}, we arrive at the following lemma. Note that our motivation is to study the pRVB state in which each hexagon is predominantly in the `-' state.

Lemma: If hexagon A is initially in a singlet state, the coupling hamiltonian $H_{AB}$ will take it to a triplet state. In particular, if A is initially in the 
the `-' state, then the coupling matrix element in Eq.\ref{eq.ABcoupling} can be non-zero only if $A_u$ - the final state - is a triplet. This is a consequence of the rotation properties of the spin operator $S_{i_b,A,\alpha}$ which acts as a triplet under rotation. 
%which dictate when $\la A_{u} \vert S_{i_b,A,\alpha} \vert A_{`-'} \ra$ can be non-zero. As the `-' state is a singlet, the spin operator can have a non-zero matrix element only when the final state is a triplet. 

We make the following assertions using this lemma. 

(i) $\la A_{-} B_{-} \vert H_{AB} \vert A_{-} B_{-} \ra =0$. Thus, the strict plaquette RVB state which has both hexagons in the pure `-' state is not preserved under the coupling Hamiltonian. In the bosonic representation, the inter-plaquette coupling does not lead to terms of the form $b_{A,-}^\dg b_{B,-}^\dg b_{A,-} 
b_{B,-}$ (equivalently, $\pbar^4$). In the condensed state, after taking inter-plaquette terms into account, there is no term proportional to $\pbar^4$ in the Hamiltonian. 

(ii) Similarly, $\la A_{-} B_{-} \vert H_{AB} \vert A_{-} B_{u} \ra =0$. There are no terms proportional to $\pbar^3$ in the Hamiltonian. 

(iii) $\la A_{u} B_v \vert H_{AB} \vert A_{-} B_{-} \ra$ can be non-zero only when $u$ and $v$ are triplet states. Furthermore, from conservation of z-component of spin, we should have $m_u + m_v =0 $. 
These matrix elements lead to $\mathcal{O}(\pbar^2)$ terms which involve creation/annihilation of pairs of triplets, i.e., we have terms of the form $\pbar^2 \Big\{ b_{i,u}^\dg b_{j,v}^\dg + b_{j,v}b_{i,u} \Big\}$.

(iv) $\la A_{-} B_v \vert H_{AB} \vert A_{u} B_{-} \ra$ can be non-zero only when $u$ and $v$ are triplet states. This gives rise to $\mathcal{O}(\pbar^2)$ terms which involve hopping of triplet operators, i.e., $\pbar^2 \Big\{ b_{j,v}^\dg  b_{i,u} \Big\} $.

(v) We can have non-zero matrix elements of the form $\la A_{-} B_{u} \vert H_{AB} \vert A_{v} B_{q} \ra$ - leading to  $\mathcal{O}(\pbar^1)$ terms. 
However, these processes are highly unlikely and can be ignored. This is justified when the pRVB order is strong and each plaquette to be predominantly in the `-' state. While the `-' boson is condensed, all other bosons will be present in very dilute concentrations. Thus, interactions between such bosons are unlikely. 

(vi) We also have terms independent of $\pbar$ which do not involve the `-' state, but these can again be ignored using the above argument.

The above statements drastically simplify the effective coupling between adjacent plaquettes A and B. Our scheme can be justified in two different ways. In the first approach, we expand the hamiltonian in powers of $\pbar$. Arguing from (i) through (iv) above, we see that the leading terms are $\mathcal{O}(\pbar^2)$. These terms are quadratic in triplet operators and involve hopping-like and pairing-like terms discussed in (iii) and (iv) respectively. We retain these leading terms alone and ignore $\mathcal{O}(\pbar), \mathcal{O}(1)$ corrections as discussed in (v), (vi) above.

A second approach is to start with the assumption of strong pRVB order. Every hexagon is predominantly occupied by the `-' boson. All other bosons are present in very dilute quantities. Thus, beginning with (v) and (vi) above, we neglect unlikely interactions between these dilute bosons. Using (i)-(iii), we find that the most likely processes are $\mathcal{O}(\pbar^2)$ terms which are quadratic in triplet operators.

The above arguments apply to any pair of nearest-neighbour plaquettes in the honeycomb lattice of Fig.\ref{fig.prvb}. They lead to an effective hamiltonian for the pRVB state which \textit{only} involves the `-' singlet and all triplet bosons. The quintets, heptets and other singlets simply drop out of the problem. 
%These states form local non-dispersing excitations which are not populated in the pRVB ground state. 
Interaction corrections arising from $\mathcal{O}(\pbar)$ terms that we have ignored can populate these states. However, we expect these corrections to be small. In the rest of this paper, we ignore the other bosons as they play no role. 

\subsection{Effective Hamiltonian for the pRVB state}

We have examined the nature of the coupling between nearest neighbour plaquettes, and deduced that they lead to hopping-like and pairing-like terms in triplet operators. There are no couplings between further neighbours as the microsopic spin Hamiltonian has only short range interactions. A cursory inspection of Fig. \ref{fig.prvb} shows that $J_1$ and $J_2$ interactions do not couple further neighbours.

The effective theory of the pRVB state can be written as 
\bea
\nn H = \sum_{i} \sum_{u=triplets} (\epsilon_{u}-\mu) b_{i,u}^\dg b_{i,u} + N\pbar^2(\epsilon_- - \mu) + N\mu \\
+\pbar^2 \sum_{\la ij \ra} \sum_{u,v = triplets}  \Big[ \left(h_{uv}^{ij}\right)  b_{i,u}^\dg b_{j,v} + \left(d_{uv}^{ij}\right) b_{i,u}^\dg b_{j,v}^\dg  + h.c.\Big].
\label{eq.effHmlt}
\eea
The first line includes the intra-plaquette terms, the index i summing over all shaded plaquettes in Fig.\ref{fig.prvb}. The chemical potential $\mu$ can be tuned to enforce single boson occupancy (on average). N denotes the number of shaded plaquettes in the system.

The second line involves inter-plaquette terms - the indices $u$ and $v$ sum over \textit{all} 27 triplet states in the single-hexagon spectrum. As argued at the end of the previous section, the quintets, heptets and singlets other than the `-' state drop out of the problem. The coefficients $\left(h_{uv}^{ij}\right)$ and $\left(d_{uv}^{ij}\right)$ are the amplitudes for triplet hopping and triplet pairing processes. The evaluation of these amplitudes is discussed in the Appendix. 

We rewrite the Hamiltonian of Eq.\ref{eq.effHmlt} by going to momentum space\cite{Fourier}. As the plaquettes form a triangular lattice, we obtain a hexagonal Brillouin zone. Fig. \ref{fig.BZs} shows the plaquette-Brillouin zone in relation to the hexagonal Brillouin zone of the underlying honeycomb lattice. 
The Hamiltonian takes the form
\bea
H\!\! =\!\!  \sum_{\bk}{}^{'}\!\! \left(\begin{array}{c}
                         b_{1,\bk}^\dg \\ \ldots \\ b_{27,\bk}^\dg \\ b_{1,-\bk} \\ \ldots \\ b_{27,-\bk}
                        \end{array}\right)^T \!\!\!
\left( \!\begin{array}{cc}
        \hat{H}_{\epsilon,\mu,h}(\bk) & \hat{H}_{d}(\bk) \\
        \hat{H}_{d}^\dg (\bk) & \hat{H}_{\epsilon,\mu,h}^T (-\bk)
       \end{array}\! \right)\!\!\!
\left(\begin{array}{c}
                         b_{1,\bk} \\ \ldots \\ b_{27,\bk} \\ b_{1,-\bk}^\dg \\ \ldots \\ b_{27,-\bk}^\dg
                        \end{array}\right).
\label{eq.tripHmlt}
%\left( \begin{array}{ccccc}
%\epsilon_{1}^t -\mu + \pbar^2 h_{\bk}^{11}  & \pbar^2 h_{\bk}^{12} & \ldots & \pbar^2 d_{\bk}^{12} & \ldots \\
%& \ddots & & & 
%\end{array}\right)
\eea
The matrix $\hat{H}_{\epsilon,\mu,h}$ involves triplet energies of the single hexagon problem, the chemical potential $\mu$ and hopping terms $h_{uv}^{ij}$. The off-diagonal blocks $\hat{H}_{d}$ involves pairing amplitudes $d_{uv}^{ij}$. The primed sum indicates that if $\bk$ is included, $-\bk$ is to excluded.

The eigenvalue spectrum of this hamiltonian can be obtained by a bosonic Bogoliubov transformation. These give the energies of `triplon' excitations above the pRVB state. The ground state energy acquires zero-point energy contributions from these triplon modes. The hamiltonian has two parameters $\pbar$ and $\mu$ which are determined self-consistently, by demanding 
\bea
\frac{\partial \la H \ra}{\partial \mu} = 0, \label{eq.mueq}\\
\frac{\partial \la H \ra}{\partial \pbar} = 0. \label{eq.pbareq}
\eea
Eq.\ref{eq.mueq} tunes the average boson density. We treat $\pbar$ as a variational parameter - Eq. \ref{eq.pbareq} minimizes the ground state with respect to $\pbar$.

For any value of $J_2 <0.5$, the hamiltonian in Eq.\ref{eq.effHmlt} can be derived and diagonalized. (For $J_2>0.5$, our plaquette operator theory is not justified as the `-' state is not the ground state of the single hexagon problem.) 
Eqs.\ref{eq.mueq}-\ref{eq.pbareq} can then be used to obtain the self-consistent solution. In the self-consistent solution, the eigenvalue of the lowest triplon mode gives the spin gap above the pRVB state.

\section{Results}
\label{sec.results}

\begin{figure}
\includegraphics[width=3.5in]{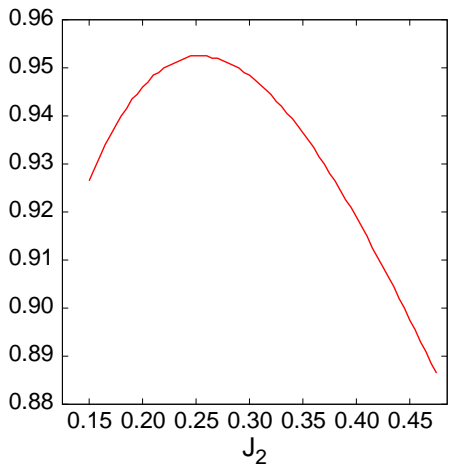} %/home/gr2/honeycombED/Draft/pbarsquared.eps}
\caption{The probability of finding a shaded plaquette in the `-' state. $\pbar^2$ as a function of $J_2/J_1$. pRVB order is strongest around $J_2\sim 0.25$. }
\label{fig.pbarsq}
\end{figure}

The plaquette operator mean-field theory is a novel method to study plaquette-ordered phases. Our plaquette operator calculation is valid in the limit of strong plaquette order, i.e., when $\pbar^2$ is close to unity - as $\pbar^2$ is the probability of finding a plaquette in the `-' state. 
As a consistency check, we plot the self-consistently obtained value of $\pbar^2$ as a function of $J_2/J_1$ in Fig.\ref{fig.pbarsq}. We find that pRVB order is strongest around $J_2 \sim 0.25$, where $\pbar^2\sim 96\%$. As seen in the figure, we find $\pbar^2 > 90\%$ for $0.15\lesssim J_2 \lesssim 0.43$. This puts our plaquette operator approach on a solid basis. 

\begin{figure}
\includegraphics[width=3.5in]{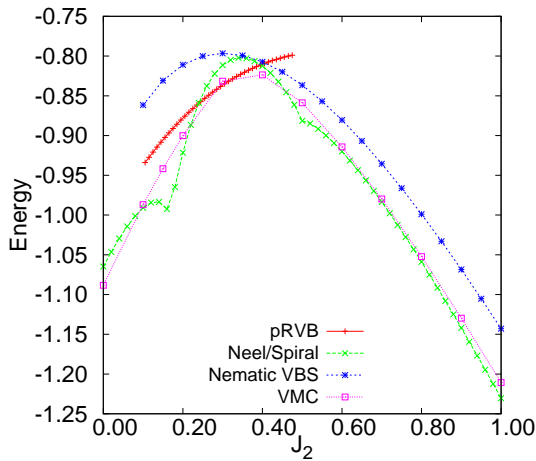} %GSenergy.eps}
\caption{The energy of the pRVB state obtained from plaquette operator theory. We compare energy per unit cell with other candidate states (see text).}
\label{fig.GSegy}
\end{figure}

In Fig.\ref{fig.GSegy}, we plot the ground state energy obtained from plaquette operator theory as a function of $J_2/J_1$. This is obtained as the zero temperature expectation value of the Hamiltonian in Eq. \ref{eq.tripHmlt}. 
To see if the pRVB state is a plausible ground state, we compare its energy with the following estimates: 

(i) N\'eel/Spiral state from large-S spin wave theory: in the semi-classical large-S limit, the ground state has N\'eel order for $J_2 < J_1/6$ and spiral order for $J_2 > J_1/6$. Linear spin wave theory gives the energy of low lying excitations\cite{Mulder}. Fig.\ref{fig.GSegy} plots the energy of the large-S ground state including zero-point contributions from spin wave excitations. To obtain the energy for the spin-1/2 case, we set $S=1/2$ in the spin wave calculation. For $0.22<\lesssim J_2 \lesssim 0.4$, the pRVB state is lower in energy than the large-S estimate.

(ii) `Nematic Valence Bond Solid': a dimer state which breaks lattice rotational symmetry has been proposed as the ground state for $J_2\gtrsim 0.4$\cite{Fouet}. This state can also be viewed as the staggered dimer state on the honeycomb lattice. We plot the energy of this state obtained using bond operator mean-field theory keeping only quadratic terms in triplet operators\cite{Mulder}. For $J_2\lesssim 0.4$, the pRVB state is lower energy than this dimer state. 

(iii) Variational Monte Carlo (VMC): recently, a VMC scheme using entangled-plaquette wavefunctions has estimated the ground state energy for the $J_1-J_2$ model\cite{Boninsegni}. Around $J_2 \sim 0.3$, our energy estimate for the pRVB state compares well with the variational result. In fact, the pRVB energy is lower than the VMC result at $J_2 =0.3$ - this may be an artifact of our mean-field theory. Corrections such as triplet-triplet interactions, coupling to quintet states, etc., may of course change the energy of the pRVB state. While the VMC study does not find pRVB order, energy comparison indicates that pRVB is a plausible ground state. 

\begin{figure}
\includegraphics[width=2.in]{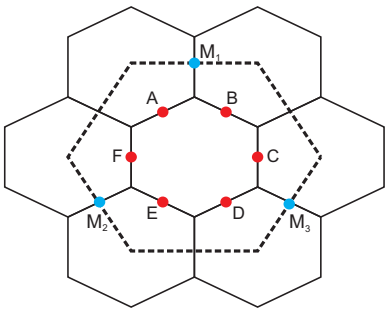}%plBZ.eps}
\caption{The plaquette Brillouin Zone (BZ): The dashed line indicates the BZ of the underlying honeycomb lattice. The pRVB state occurs on a triangular lattice of plaquettes - the corresponding BZ is shown by solid lines using the extended zone scheme. The M points of the plaquette-triangular BZ are highlighted by blue/red circles - these are the wavevectors at which the lowest triplet excitations occur.}
\label{fig.BZs}
\end{figure}

Comparing the energy of the pRVB state with the above candidates, we conclude that the ground state may have pRVB order for $0.25 \lesssim J_2 \lesssim 0.4$. Next, we plot the spin gap as a function of $J_2/J_1$ - this is the energy of the lowest triplon state. We find that this lowest triplon state always occurs at the M points of the plaquette-Brillouin zone (see Fig.\ref{fig.BZs}). The spin gap reaches a maximum value of $\sim 0.7 J_1$ at $J_2 \sim 0.225$.
Surprisingly, the spin gap does not close as $J_2$ is varied. We discuss its implications in the next section.

\begin{figure}
\includegraphics[width=3.5in]{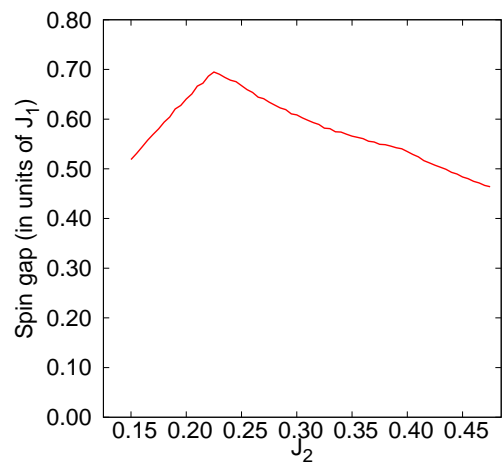}%{/home/gr2/honeycombED/Draft/spingap.eps}
\caption{The spin gap obtained from plaquette operator theory as a function of $J_2$. The gap is in units where $J_1$ is unity.}
\label{fig.spingap}
\end{figure}

\section{Discussion}
\label{sec.discussion}

We have developed an effective theory for the proposed pRVB state in the spin-1/2 $J_1-J_2$ model on the honeycomb lattice. Using plaquette-operator mean-field theory, we have calculated the properties of this state and the spectrum of low-lying excitations above it. 

An important outcome of our calculation is to clarify the nature of the pRVB state. The low energy state is composed of a $\sqrt{3}\times\sqrt{3}$ ordering of plaquettes in the `-' state. We begin our calculation from the limit of decoupled hexagons, wherein the `-' state is clearly lower in energy for $J_2 < 0.5$. After including inter-hexagon coupling, we get a consistent theory of the pRVB state. Thus, our plaquette wavefunction changes sign upon rotation by 60$^\circ$. Under rotation by angle $\phi$, the plaquette wavefunction transforms as $e^{i3\phi}$ - having f-wave symmetry. 

As we argue below, we have ruled out s-wave pRVB order involving condensation of the `+' state. In the limit of disconnected hexagons, the `+' state is an excited state. As such, we expect s-wave pRVB order to be unstable for this reason. However, strong inter-hexagon coupling could possibly stabilize this phase by means of a large kinetic gain from delocalizing the triplet excitations. To rule out this possibility, we redid the plaquette operator calculation for the s-wave case. We do not find a consistent s-wave pRVB solution, thus ruling out strong s-wave ordering.

The earliest study of plaquette order on the honeycomb lattice describes s-wave plaquette order in a quantum dimer model. The plaquette wavefunction is \textit{symmetric} under rotation\cite{Moessner}. However, for the $J_1-J_2$ model at hand, Ref.\onlinecite{Albuquerque} indicates that the antisymmetric combination is favoured - the authors refer to this order as d-wave, while it is more precisely called f-wave. We reaffirm this f-wave nature using our plaquette operator approach.

\subsection{Excitations of pRVB state}
\label{ssec.excitations}

This paper presents the first systematic study of the pRVB state and its excitations. As shown in Fig.\ref{fig.spingap}, the spin gap does not close within our theory. And the lowest lying triplet modes occur at the M points of the plaquette-BZ. In terms of the underlying honeycomb lattice, the lowest triplet state will  occur at wavevectors shown in fig.\ref{fig.BZs}. Assuming that pRVB order occurs in the $J_1-J_2$ model, numerical methods such as ED or DMRG can see this by calculating the spin-spin structure factor in the lowest triplet state. The structure factor should show peaks at the wavevectors shown in Fig.\ref{fig.BZs}.
Indeed, this can serve as a test for pRVB order.  

As is familiar from the study of dimer states\cite{Nikuni}, an applied magnetic field will lower the energy of triplet modes. At some critical field, the lowest triplet mode(s) will become gapless and undergo Bose condensation giving rise to magnetic order. In our case, the lowest triplet modes are triply degenerate as they occur at the three M points of the plaquette-BZ. Beyond the critical field, each mode can condense with an independent amplitude and phase leading to six degrees of freedom. This makes it difficult to predict the nature of the magnetic ordering beyond the critical field. At the level of these arguments, we can rule out field-induced N\'eel ordering as the lowest triplets do not occur at the $\Gamma$ point. This picture may change once we account for defects in pRVB order as discussed below.

\subsection{Phase transitions}
\label{ssec.phasetransitions}

For the case of dimer-N\'eel transitions, bond operator theory is remarkably sucessful in describing the nature of the phase transition 
%(for example, Quantum Monte Carlo results\cite{Sandvik} agree with bond operator estimates\cite{Matsushita} for the critical for dimer-N\'eel transitions in bilayer systems).
Starting from the dimer state, bond operator theory provides a consistent theory of spin-1 triplon excitations which obey bosonic statistics. The transition to N\'eel order occurs via Bose-condensation of triplons when the spin gap closes\cite{SachdevBhatt, Ruegg}. 

In our honeycomb lattice model, our plaquette operator theory gives a consistent description of the pRVB state which is most stable around $J_2\sim 0.25$. 
ED calculations suggest that a pRVB state is sandwiched between a N\'eel state(for $J_2\lesssim 0.2)$ and a magnetic phase with stripe order (for $J_2 \gtrsim 0.4$). While the phase diagram definitely contains N\'eel order, the existence of a stripe magnetic phase is still open to question. It has been suggested that the ground state for $J_2 \gtrsim 0.4$ has dimer order which breaks lattice rotational symmetry\cite{Fouet,Mulder}.  

Starting from the pRVB state, as $J_2$ is decreased, we expect to see a N\'eel transition near $J_2 \sim 0.2$. Na\"ively, we may expect such a transition to be driven by condensation of triplons giving rise to a continuous Quantum Phase Transition (QPT) in the same universality class as the BEC transition in 2+1 dimensions. 
However, the spin gap obtained from plaquette operator theory does not close as $J_2$ is lowered, thus ruling out a BEC transition. 
This leaves us with two possibilities concerning the pRVB-N\'eel transition: (i) a first order phase transtion, or (ii) a deconfined QPT mediated by topological defects in pRVB order. 
In terms of conventional Landau theory, we generically expect a first order QPT as the N\'eel state and pRVB state break different symmetries. This is always a possibility and can only be checked using precise numerical techniques. The second possibility, a deconfined QPT\cite{deconfined,LevinSenthil}, is a very exciting prospect. So far, such a transition has not been observed in simple models involving Heisenberg-like couplings.

A deconfined QPT would be driven by proliferation of defects in the pRVB order parameter - in this case, Z$_3$ vortices. As pointed out in Ref.\cite{CenkeXu}, the core of such a vortex will necessarily bind a dangling spin or a `spinon'. The critical theory of a continuous pRVB-N\'eel transition cannot be cast in terms of Landau-Ginzburg order parameters, but in terms of these spinful defects. We expect the theory to be very similar to the one outlined in Ref.\cite{LevinSenthil} - at the critical point, the vortices become gapless and the Z$_3$ anisotropy may become irrelevant. 

Starting from the pRVB state, when we increase $J_2$, there must be a transition into a state which breaks lattice rotational symmetry - a nematic Valence Bond Solid\cite{Mulder} or a magnetic phase with stripe order\cite{Albuquerque}. The spin gap does not close in our plaquette operator theory. As suggested by ED results\cite{Albuquerque}, we surmise that this is a first order transition. Future numerical studies will be able to test this notion. 

\acknowledgements
We thank I. Rousochatzakis for many helpful discussions. 

\appendix
\section{Matrix elements}

\begin{figure}
\includegraphics[width=2.5in]{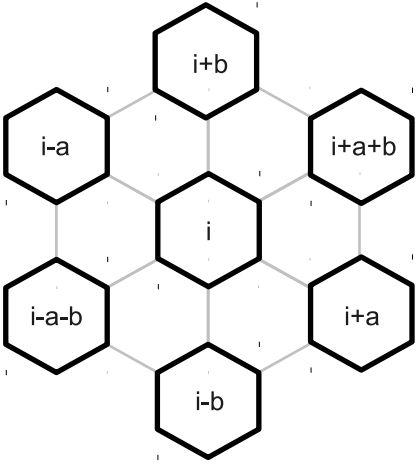}
\caption{A plaquette i and its nearest-neighbours. $\hat{a}$ and $\hat{b}$ are the primitive vectors of the $\sqrt{3}\times\sqrt{3}$ lattice of plaquettes. }
\label{fig.nnbrs}
\end{figure}

Fig. \ref{fig.nnbrs} shows a plaquette i and its six nearest neighbours. Let us first consider the nearest neighbour pair $(i,i-\hat{b})$. This pair is identical to that shown in Fig.\ref{fig.twohexagons} if we identify $(i)$ as A and $(i-\hat{b})$ as B. Inter-plaquette coupling leads to hopping-like and pairing-like terms between triplon operators on the two sites. Below, we evaluate the coefficients $h_{uv}^{i,i-\hat{b}}$ and $d_{uv}^{i,i-\hat{b}}$.

The generic inter-plaquette interaction process is indicated in Eq. \ref{eq.ABcoupling}. The hopping-like term arises the following matrix element.
\bea
h_{uv}^{i,i-\hat{b}} = \la (i)_u (i-\hat{b})_- \vert H_{i,i-\hat{b}} \vert (i)_- (i-\hat{b})_v \ra.
\label{eq.hAB}
\eea
This matrix element can be evaluated numerically using two pieces of information: (i) the explicit eigenstates of the single hexagon problem and (ii) the explicit form of the inter-plaquette coupling given in Eq. \ref{eq.ABcoupling}.
In plaquette operator notation, this matrix element leads to the process $b_{i,u}^\dg b_{i-\hat{b},-}^\dg b_{i,-} b_{i-\hat{b},v}$. As the `-' bosons are condensed with amplitude $\pbar$, we rewrite this as
\bea
H_{hopping} \sim h_{uv}^{i,i-\hat{b}} \pbar^2 b_{i,u}^\dg b_{i-\hat{b},v}.
\eea
Similarly, the coefficient of the pairing term is given by 
\bea
d_{uv}^{i,i-\hat{b}} = \la (i)_u (i-\hat{b})_v \vert H_{i,i-\hat{b}} \vert (i)_- (i-\hat{b})_- \ra.
\eea
This gives rise to $b_{i,u}^\dg b_{i-\hat{b},v}^\dg b_{i,-} b_{i-\hat{b},-}$, which can be rewritten as 
\bea
H_{pairing} \sim d_{uv}^{i,i-\hat{b}} \pbar^2 b_{i,u}^\dg b_{i-\hat{b},v}^\dg.
\eea

Having evaluated these terms for one nearest neighbour pair, we can evaluate the matrix elements for all other pairs using the rotation operator.
Let us consider three nearest neighbour pairs in Fig. \ref{fig.nnbrs}: $(i,i-\hat{b})$, $(i,i+\hat{a})$ and $(i,i+\hat{a}+\hat{b})$. Clearly, these three pairs are related by $\hat{R}$, rotation by 60$^\circ$, about the center of the hexagon $i$. Formally, in terms of the inter-plaquette couplings, we may write 
\bea
H_{i,i+\hat{a}} = \hat{R} H_{i,i-\hat{b}} \hat{R}^{\dg}.
\eea
The rotation operator $\hat{R}$ maps the plaquette $(i-\hat{b})$ to $(i+\hat{a})$. In addition, it rotates the hexagons by 60$^\circ$.

Using this relation, we can obtain the hopping matrix element $h_{uv}^{i,i+\hat{a}}$:
\bea
\nn h_{uv}^{i,i+\hat{a}} = \la (i)_u (i+\hat{a})_- \vert H_{i,i+\hat{a}} \vert (i)_- (i+\hat{a})_v \ra \\
\nn =  \la (i)_u (i+\hat{a})_- \vert \hat{R} H_{i,i-\hat{b}} \hat{R}^{\dg} \vert (i)_- (i+\hat{a})_v \ra \\
\nn =  r_{u} r_v^* \la (i)_u (i-\hat{b})_- \vert H_{i,i-\hat{b}} \vert (i)_- (i-\hat{b})_v \ra \\
= r_{u} r_v^* h_{uv}^{i,i-\hat{b}},
\eea
where $r_u$ and $r_v$ are the rotation-eigenvalues of the single-hexagon states $\vert u \rangle$  and $\vert v \rangle$ respectively. 

The coefficient of the pairing-like term can be found likewise:
\bea
\nn d_{uv}^{i,i+\hat{a}} = \la (i)_u (i+\hat{a})_v \vert H_{i,i+\hat{a}} \vert (i)_- (i+\hat{a})_- \ra \\
\nn =  \la (i)_u (i+\hat{a})_v \vert \hat{R} H_{i,i-\hat{b}} \hat{R}^{\dg}\vert (i)_- (i+\hat{a})_- \ra \\
\nn =  r_{u} r_v (r_-^*)^2 \la (i)_u (i-\hat{b})_v \vert H_{i,i-\hat{b}} \vert (i)_- (i-\hat{b})_- \ra \\
= r_{u} r_v (r_-^*)^2 d_{uv}^{i,i-\hat{b}}.
\eea

Similarly, by applying one more rotation, we can obtain
\bea
\nn h_{uv}^{i,i+\hat{a}+\hat{b}} = r_{u}^2 (r_v^*)^2 h_{uv}^{i,i-\hat{b}},
d_{uv}^{i,i+\hat{a}+\hat{b}} = r_{u}^2 r_v^2 (r_-^*)^4 d_{uv}^{i,i-\hat{b}}.
\eea


\begin{thebibliography}{}
\bibitem{Fourier}
We use the transform convention $b_{i,u} \equiv \frac{1}{\sqrt{N}}\sum_{\bk} b_{\bk,u}e^{i\bk \cdot \br_i}$. The wavevectors $\bk$ lie in the plaquette-Brillouin zone shown in Fig.\ref{fig.BZs}.
\end{thebibliography}
\end{document}